
\magnification 1200

\def\dsp{\baselineskip=22pt plus 1pt minus 1pt}
\def\rf{\hfill\break}

\def\ppp{\par \smallskip \noindent \hangindent .5in \hangafter 1}
\def\e{\eta}

\def\g{\gamma}
\def\G{\bar \gamma}

\def\p{\partial}

\def\cl{\centerline}
\def\sles{\lower2pt\hbox{$\buildrel {\scriptstyle <}
   \over {\scriptstyle\sim}$}}
\def\sgreat{\lower2pt\hbox{$\buildrel {\scriptstyle >}
   \over {\scriptstyle\sim}$}}
\def\sg{r^2}
\rf
\dsp
\rf
\bigskip
\centerline{\bf HYDRODYNAMICS OF RELATIVISTIC FIREBALLS}
\bigskip
\bigskip
\cl{Tsvi Piran,$^{1,2}$ Amotz Shemi$^3$ and Ramesh Narayan$^1$}
\bigskip
\item{1.} Harvard-Smithsonian Center for Astrophysics, Cambridge, MA 02138,
USA.
\item{2.} Racah Institute for
Physics, The Hebrew University, Jerusalem, Israel.
\item{3.} Department of Physics and Astronomy, Tel Aviv University, Tel Aviv,
Israel.
\bigskip
\bigskip
\centerline{\bf Summary}
\bigskip
We examine analytically and numerically the evolution of a
relativistic fireball.  We show that, after an early rearrangement
phase, most of the matter and energy in the fireball is concentrated
within a narrow shell.  The shell propagates at nearly the speed of
light, with a frozen radial profile and according to a simple set of
scaling laws.  The spectrum of the escaping radiation is harder at
early times and softer later on.  The results are applicable to models
of $\gamma$-ray bursts.
\vfill\eject

\bigskip
\centerline{\bf 1. Introduction}
\bigskip

The sudden release of a large quantity of gamma ray photons into a
compact region can lead to an opaque photon--lepton ``fireball''
through the production of electron--positron pairs.  The term
``fireball" refers here to an opaque radiation--plasma whose initial
energy is significantly greater than its rest mass.  The formation and
evolution of fireballs is of interest in astrophysics (Cavallo \& Rees
1978), especially for the understanding of gamma-ray bursts at
cosmological distances (Goodman 1986, Paczynski 1986, 1990, Shemi \& Piran
1990, Narayan, Paczy\'nski \& Piran 1992, M\'es\'zaros and Rees
1992a,b), or in the halo of the Galaxy (Piran \& Shemi 1993).

In this paper we investigate the hydrodynamics of fireballs.  We begin
by summarizing in this section several qualitative results that are
already known in this problem.  Because of the opacity due to pairs,
the radiation in a fireball cannot initially escape.  Instead, the
fireball expands and cools rapidly until the temperature drops below
the pair--production threshold and the plasma becomes transparent.  In
addition to radiation and $e^+e^-$ pairs, astrophysical fireballs may
also include some baryonic matter which may be injected with the
original radiation or may be present in an atmosphere surrounding the
initial explosion.  The electrons associated with this matter increase
the opacity, delaying the escape of radiation.  More importantly, the
baryons are accelerated with the rest of the fireball and convert part
of the radiation energy into bulk kinetic energy.

As the fireball evolves two important transitions take place.  One
transition corresponds to the change from optically thick to optically
thin conditions.  As long as the total opacity (pairs + matter) is
large the plasma expands adiabatically as a perfect fluid (Goodman
1986).  However, once $\tau$ drops below 1, the photons and baryons
decouple from each other and continue their evolution independently
and without interaction.  The second transition corresponds to the
switch from radiation dominated to matter dominated conditions, i.e
from $\e >1$ to $\e <1$, where $\e$ is the ratio of the radiation
energy $E$ to the rest energy $M$: $\e \equiv E/Mc^2$ (Cavallo \& Rees
1978, Shemi \& Piran 1990).  In the early radiation dominated stages
when $\e>1$, the fluid accelerates in the process of expansion,
reaching relativistic velocities and large Lorentz factors.  The
kinetic energy too increases proportionately.  However, later when $\e
< 1$, the fireball becomes matter dominated and the kinetic energy is
comparable to the total initial energy.  The fluid therefore coasts
with a constant radial speed.  The overall outcome of the evolution of
a fireball then depends critically on the value of $\e$ when $\tau$
reaches unity.  If $\e>1$ when $\tau=1$ most of the energy comes out
as high energy radiation, whereas if $ \e < 1 $ at this stage most of
the energy has already been converted into kinetic energy of the
baryons.

The opacity itself has a contribution from electron-positron pairs as
well as electrons associated with the baryons.  Initially, when the
local temperature $T$ is large, the opacity is dominated by $e^+e^-$
pairs (Goodman, 1986) .  But this opacity, $\tau_p$, decreases
exponentially with decreasing temperature, and falls to unity when
$T=T_p\approx 20$ KeV.  The matter opacity, $\tau_b$, on the other
hand decreases only as $R^{-2}$, where $R$ is the radius of the
fireball.  If at the point where $\tau _p =1$, $\tau_{b}$ is still $ >
1$, then the final transition to $\tau=1$ is delayed and occurs at a
cooler temperature.

The initial ratio of radiation energy to mass, $\e_i$, determines in
what order the above transitions take place.  Shemi and
Piran (1990) identified four regimes: (i) $\e_i >
\e_{pair} = [3 \sigma_T^2 E_i \sigma T_p^4 /4 \pi m_p^2 c^4 R_i
]^{1/2}$ (where $E_i$ and $R_i$ are the initial energy and radius): In
this regime the effect of the baryons is negligible and the evolution
is of a pure photon-lepton fireball.  When the temperature reaches
$T_p$, the pair opacity $\tau_p$ drops to 1 and $\tau _b \ll 1$.  At
this point the fireball is radiation dominated ($\e >1$) and so most
of the energy escapes as radiation.  (ii) $\e_{pair} > \e_i > \e_{b} =
(3 \sigma_T E_i / 8 \pi m_p c^2 R_i^2)^{1/3} $: Here, in the late
stages, the opacity is dominated by free electrons associated with the
baryons.  The comoving temperature therefore decreases far below $T_p$
before $\tau$ reaches unity.  However, the fireball continues to be
radiation dominated as in the previous case, and most of the energy
still escapes as radiation.  (iii) $\e_{b} >
\e_i > 1$: The fireball becomes matter dominated before it becomes
optically thin.  Therefore, most of the initial energy is converted
into bulk kinetic energy of the baryons, with a final Lorentz factor
$\G_f = \e_i+1$.  (iv) $\e_i < 1$: This is the Newtonian regime.  The
rest energy exceeds the radiation energy and the expansion never
becomes relativistic.

The above summary describes the qualitative features of a roughly
homogeneous expanding fireball.  In this paper we investigate some
aspects of the evolution of inhomogeneous fireballs with a non-uniform
radial profile.  We show in \S 2 that, after an initial rearrangement
phase, the evolution is well described by an asymptotic large $\g$
solution.  The radial profile of the fireball remains frozen over most
of this phase, and each streamline follows simple scaling laws as a
function of radius.  In \S 3 we solve numerically the adiabatic
expansion of a spherical fireball and compare the results with the
asymptotic solution.  We show that the agreement with the theoretical
solution is good.  Finally in
\S 4 we summarize the results and discuss their implications.

\bigskip
\centerline{\bf 2. Scaling Laws and Asymptotic Solutions}
\bigskip

We consider a spherical fireball with an arbitrary radial distribution
of radiation and matter.  Under optically thick conditions there is
strong coupling between the photons and baryons, and so
the radiation and matter at each radius behave like a single fluid,
moving with the
same velocity. Since the radiation pressure dominates, the pressure
$p$ and the energy density $e$ are related by $p =e/3 $, and we
can rewrite the standard relativistic conservation equations of baryon
number and energy momentum (Weinberg, 1973) as:
$$ {\p \over \p
t}( n \g ) + {1 \over \sg} {\p \over \p r}(\sg n u) = 0,
\eqno(1)
$$
$$
{\p \over \p t}( e^{3/4} \g ) +
{1 \over \sg}
{\p \over \p r}(\sg e^{3/4} u ) = 0,
\eqno (2)
$$
$$
{\p \over \p t}\left [ \left (n + {4\over 3} e\right ) \g u \right ] +
{1 \over \sg}
{\p \over \p r}\left [ \sg \left (n + {4 \over 3} e\right ) u^2 \right ] =
- {1 \over 3}{\p e \over \p r},
\eqno(3)
$$
where $\g = u^t$, $u=u^r = \sqrt{\g^2-1}$, and we use units in
which $c=1$ and the mass of the particles $m=1$.
The mass density,
$n$, the total energy density, $e$ (which includes contributions from the
radiation as well as the relativistic
electron-positron pairs at temperatures where the latter are present),
and the pressure, $p$, are measured in
the local frame of the fluid, but $r$ and $t$ are in the observer
frame.

Change variables from $r,~t$ to $r,~s=t-r$.  Equations (1)--(3) then
become
$$
{1\over r^2}{\p\over \p r}(r^2nu)=-{\p\over\p s}\left
({n\over \g+u} \right ),\eqno (4)
$$
$$ {1\over r^2}{\p\over \p
r}(r^2e^{3/4}u)=-{\p\over\p s}\left ({e^{3/4} \over \g+u}\right
),\eqno (5)
$$
$$ {1\over r^2}{\p\over \p r}\left [r^2\left
(n+{4\over 3}e\right )u^2\right ]= -{\p\over\p s}\left [\left
(n+{4\over 3}e\right ) {u\over \g +u}\right ] +{1\over 3}\left [{\p
e\over \p s} -{\p e\over \p r}\right ],\eqno (6)
$$
where the derivative $\p /\p r$ now refers to constant $s$, i.e.
is calculated along a characteristic moving outward at the speed of
light.  After a short acceleration phase we expect that the motion of
a fluid shell will become highly relativistic ($\g \gg 1$).  If we
restrict our attention to the evolution of the fireball from this
point on, we may treat $\g ^{-1}$ as a small parameter and set $\g
\approx u$, which is good to order $o(\g ^{-2})$.  Then, under a wide
range of conditions, which we discuss below, the quantities on the
right-hand sides of equations (4)--(6) are significantly smaller
than those on the left.  We therefore set the terms on the right
to zero, and obtain the following conservation laws for each fluid  shell:
$$
r^2n\g = {\rm constant}, \qquad r^2e^{3/4}\g = {\rm constant},
\qquad r^2\left (n+{4\over 3}e\right )\g ^2 = {\rm constant}.\eqno (7)
$$

Two regimes of behavior are then immediately apparent.  In the
radiation-dominated phase ($e\gg n$), we have
$$
\g\propto r,\qquad n\propto r^{-3},\qquad e\propto r^{-4},
\qquad T_{obs}\sim {\rm constant},\eqno (8)
$$
where $T_{obs}\propto \g e^{1/4}$ is the temperature of the radiation
as seen by an observer at infinity.  (Strictly, the radiation
temperature depends on $e_r$, the energy density of the photon field
alone; for $T \ll m_ec^2$, $e_r=e$, but for $T> m_e c^2$, $e$ contains
an additional contribution from the electron position pairs, see Shemi
\& Piran 1990; we neglect this complication for simplicity). The
scalings of $n$ and $e$ given in (8) correspond to those of a fluid
expanding uniformly in the comoving frame.  Indeed, all four scalings
in eq (8) were derived for a homogeneous radiation dominated fireball
by Shemi \& Piran (1991, see also Goodman 1986) by noting the analogy
with an expanding universe.  What we have shown here is that the same
relations are valid for each individual radial shell
in the fireball even in the more general
inhomogeneous case.  In fact, these scaling laws also apply to
Paczy\'nski's (1986) solution for a steady state relativistic wind.
When we neglect the right hand sides of eqs. (4)--(6) the problem
becomes effectively only $r$ dependent.

Although the fluid is approximately homogeneous in its own frame,
because of Lorentz contraction it appears as a narrow shell in the
observer frame, with a radial width given by $\Delta r \sim r/\g \sim
{\rm constant}\sim R_i$, where $R_i$ is the initial radius of the
fireball.  We can now go back to eqs (4)--(6) and set $\p /\p s \sim
\g /r$.  We then find that the terms we neglected on the right hand
sides of these equations are smaller than the terms on the left by a
factor $\sim 1/\g$.  Therefore, the conservation laws (7) and the
scalings (8) are valid so long as the radiation-dominated fireball
expands ultra-relativistically with large $\g$.  The only possible
exception is in the very outermost layers of the fireball where the
pressure gradient may be extremely steep and $\p /\p s$ may be $\gg \g
/r$.  Ignoring this minor deviation, we interpret eq (7) and the
constancy of the radial width $\Delta r$ in the observer frame to mean
that the fireball behaves like a pulse of energy with a frozen radial
profile, accelerating outward at almost the speed of light.

In the alternate matter-dominated regime ($e\ll n$), we obtain from
eq (7) the following different set of scalings,
$$
\g\rightarrow {\rm constant},\qquad n\propto r^{-2},
\qquad e\propto r^{-8/3},\qquad T_{obs}\propto r^{-2/3}.\eqno (9)
$$
The modified scalings of $n$ and $e$ arise because the fireball now
moves with a constant radial width in the comoving frame.  (The
steeper fall-off of $e$ with $r$ is because of the
work done by the radiation through tangential expansion.)  Moreover,
since $e\ll n$, the radiation has no important dynamical effect on the
motion and produces no significant radial acceleration.  Therefore,
$\g$ remains constant on streamlines and the fluid coasts with a constant
asymptotic
radial velocity.  Of course, since each shell moves with a
velocity that is slightly less than $c$ and that is different from
one shell to the next, the frozen pulse
approximation on which eq (7) is based must ultimately break down at
some large radius.  We consider this question below, but first
continue with our investigation of the approximate relations in eq (7).

A scaling solution that is valid in both the radiation-dominated and
matter-dominated regimes, as well as in the transition zone in
between, can be obtained by combining the conserved quantities in eq
(7) appropriately.  Let $t_0$ be the time and $r_0$ be the radius at
which a fluid shell in the fireball first becomes ultra-relativistic,
with $\g \ \sgreat\ {\rm few}$.  Label various properties of the shell at
this time by a subscript $0$, e.g. $\g _0$, $n_0$, $e_0$, and define
$\e = e/n$, $\eta _0 = e_0/n_0$.
Defining the auxiliary quantity $D$, where
$$
{1 \over D} \equiv
{\g_0 \over \g } + {3\g_0 \over 4\eta_0 \g } - {3\over 4\eta_0},\eqno(10)
$$
we find that
$$
r = r_0 { \g_0^{1/2} D^{3/2} \over \g^{1/2}},
\qquad n = {n_0 \over D^3},
\qquad e = {e_0 \over D^4},
\qquad \e = {\e _0\over D}.\eqno (11)
$$
These are parametric relations which give $r,~n,~e$, and $\e$ of each fluid
shell at any time in terms
of the $\g$ of the shell at that time.  The relation
for $r$ in terms of $\g$ is a cubic equation.  This can in principle be
inverted to yield $\g(r)$, and thereby $n,~e$, and $\e$ may also be expressed
in terms of $r$.

The parametric solution (11) describes both the radiation-dominated
and matter-dominated phases of the fireball within the frozen pulse
approximation.  For $\g\ll \e _0\g _0$, the first term in eq (10)
dominates and we find $D\propto r$, $\g\propto r$, recovering the
radiation-dominated scalings of eq (8).  This regime extends out to a
radius $r\sim \e_0 r_0$.  At larger radii, the first and last terms in
(10) become comparable and $\g$ tends to its asymptotic value of $\g_f
= (4\e_0/3+1)\g_0$.  This is the matter dominated regime.  (The
transition occurs when $4e/3=n$, which happens when $\g=\g_f/2$.)  In
this regime, $D\propto r^{2/3}$, leading to the scalings in eq (9).

Ultimately, all the energy in the fireball is concentrated in the
kinetic energy of the matter, and this determines the value of $\g_f$.
Interestingly, if we write $\g_f$ in terms of the initial parameters
of the fireball at time $t=0$, we find $\g_f =
\e _i+1$, whereas, when we write it in terms of $\e_0$, $\g_0$, we have
the additional factor of $4/3$ as written above.  Both formulae
represent energy conservation, but the component $T^{tt}$ of the
energy momentum tensor behaves differently in the two cases.  At time
$t=0$, the fluid is at rest and the radiation energy density is merely
$e$, whereas at $t=t_0$, the fluid is already moving highly
relativistically and there is an additional contribution to the energy
from the moving pressure, $T^{tt}=\g_0^2e+(1/3)u_0^2e\sim
(4/3)\g_0^2e$.

Let us now return to a consideration of very late times in the
matter-dominated phase when the frozen pulse approximation begins to
break down.  We have already seen that in this phase the radiation
density $e$ is much smaller than the matter density $n$, and also that
$\g$ tends to a constant value $\g_f$ for each shell.  We may
therefore neglect the term $-(1/3)(\p e /\p r)$ in eq (3) and treat
$\g$ and $u$ in eqs (1)--(3) as constants.  We then find that the flow moves
strictly along the characteristic, $\beta _f t-r={\rm constant}$, so
that each fluid shell coasts at a constant radial speed, $\beta
_f=u_f/\g_f$.  Let us label the baryonic shells in the fireball by a
Lagrangian coordinate $R$, moving with a fixed Lorentz factor
$\g_f(R)$, and let $t_c$ and $r_c$ represent the time and radius at
which the coasting phase begins, which corresponds essentially to the
point at which the fluid makes the transition from being radiation
dominated to matter dominated.  We then find
$$
r(t,R) - r_c(R) = {\sqrt{\g_f^2(R)-1} \over \g_f(R)} (t-t_c(R))
\approx \left [ 1-{1 \over 2 \g_f^2(R)}\right ] [t-t_c(r)].\eqno(12)
$$
The separation between two neighboring shells
separated by a Lagrangian distance $\Delta R$ varies during the
coasting phase as
$$
\left [{d (\partial r /\partial R) \over dt}\right ] \Delta R =
\left [{ 1 \over
\gamma_f(R)^3} {\partial \gamma_f \over \partial R}\right ] \Delta R.\eqno
(13)
$$
Thus the width of the pulse at time $t$ is $\Delta r(t) \approx
\Delta r_c + \Delta \gamma_f (t-t_c) / \bar\gamma_f^3
\approx R_i + (t-t_c) /\G_f^2 $, where $\Delta r_c\sim R_i $ is the width of
the fireball when it begins coasting, $\bar\g_f$ is the mean $\g_f$ in
the pulse, and $\Delta \gamma_f\sim \bar\g_f$ is the spread of $\g_f$
across the pulse.  From this result we see that there are two separate
regimes in the fireball evolution even within the matter dominated
coasting phase.  So long as $t-t_c < \bar\g_f^2R_i$, we have a
frozen-coasting phase in which $\Delta r$ is approximately constant
and the frozen pulse approximation is valid.  In this regime the
scalings in eq (9) are satisfied.  However, when $t-t_c >\bar\g_f^2
R_i$, the fireball switches to an expanding-coasting phase where
$\Delta r \propto t-t_c$ and the pulse width grows linearly with time.
In this regime the scaling of $n$ reverts to $n\propto r^{-3}$, and,
if the radiation is still coupled to the matter, $e\propto r^{-4}$.

Independently of the above considerations, at some point during the
expansion, the fireball will become optically thin and the radiation
will decouple.  From this stage on the radiation and the baryons no
longer move with the same velocity and the radiation pressure vanishes,
leading to a breakdown of equations (2) and (3).  The radiation will
now coast with a speed exactly equal to $c$ and with a constant radial
width.  The radiation energy density will clearly scale as $e\propto
r^{-2}$.  The baryon shells on the other hand will coast with their
own individual velocities.  If the fireball is already in the matter
dominated coasting phase there will be no change in the propagation of
the baryons. However, if the fireball is in the radiation dominated
phase when it becomes optically thin, then the baryons will switch
immediately to a coasting phase.

\bigskip
\centerline{\bf 3. Numerical Results}
\bigskip

We have developed a spherically symmetric relativistic code that
follows the evolution of a mass-loaded fireball from an initial
configuration at rest via the acceleration phase into the asymptotic
frozen pulse regime.  The code is Eulerian and employs a second order
conserved scheme (Bowers and Wilson, 1991) modified to take care of
the extreme relativistic Lorentz factors encountered in this problem.
Our scheme is quite different from the one used by Vitello and Salvati
(1976), who studied a similar problem.  The code has passed several
standard tests, including the Richardson test, i.e. the results
converge satisfactorily as the grid size is decreased.

The initial profiles for the cases that we present here are:
$$
e(r,t=0) = {e_0 \over (R_i^8 + r^8)},\qquad n(r,0)={ e(r,0) \over
\eta_i},\qquad \g (r,0) = 1, \eqno(14)
$$
where we choose $R_i$, the initial width of the fireball, to be unity.
The initial radiation density $e_0$ is in arbitrary units, and we
assume a constant ratio $\e_i$ of energy to mass. The energy density
falls off sufficiently rapidly with radius in the initial profile that the
external density is negligible compared to the interior density.
However, we cannot set the exterior density exactly to zero since this
leads to numerical problems.  We have explored different initial
conditions, and find different pulse shapes, but the overall
qualitative behavior is generally similar to the results described
below.

Fig. 1.a shows a sequence of profiles in the observer frame of the
energy density, $\g e$, and the mass density, $\g n$, for a simulation
with $\eta_i=50$.  Three phases of evolution are apparent, viz. an
initial acceleration/rearrangement phase, a short radiation dominated
phase, and a final matter dominated phase.  Conservation of energy
requires that, asymptotically, the average Lorentz factor of the
expanding fluid should be $\bar\g _f =\eta_i+1$.  We compute $\bar\g$
at each stage by means of
$$
\bar \g = {\int T^{tt} r^2 dr \over
\sqrt{ (\int T^{tt} r^2 dr)^2 - (\int T^{tr} r^2 dr )^2 } }
= {\int T^{tt} r^2 dr \over
\sqrt{[ \int (T^{tt} + T^{tr}) r^2 dr  ]
[\int (T^{tt} -  T^{tr}) r^2 dr] } } , \eqno(15)
$$
where the second expression is preferable for numerical accuracy.
The average Lorentz factor in this simulation does
approach the expected asymptotic value, but it does not quite reach it
because we did not let the computation continue for a long enough time.

Early in the evolution, the pulse rearranges itself during a brief
acceleration phase.  Although with our choice of initial data
the ratio between the energy and mass density, $\e$, begins with a
constant value throughout the fireball (eq 14), $\e$ changes during this
phase and no longer remains constant within the pulse. Generally,
$\eta$ ends up being smaller in the inner parts of the fireball and
larger on the outside.  This can be seen in Fig. 2 which shows $n$,
$e$ and $\gamma$ at a fixed moment of time after the end of the early
phase of rearrangement. After this phase, the shape of the pulse is
frozen and the fireball evolves through a radiation-dominated phase to
a matter-dominated state.  The transition from radiation dominated to
matter dominated can be clearly seen in Fig. 1.a where the matter
density is initially lower than the energy density but becomes larger
by the end of the computation.

The profile of the Lorentz factor $\g $ at various times in the pulse
is shown in Fig. 1.b.  We see that $\g$ varies significantly across
the pulse.  Whereas the mean $\bar\g_f$ cannot exceed the asymptotic
value of $\e_i+1$, the maximum Lorentz factor within the pulse is
larger and in fact increases throughout the evolution.  This
happens as the outermost layers of the fireball keep accelerating, as
can be seen in Fig. 2.  The Lorentz factor therefore peaks ahead of
the energy density in a low density region, and a small fraction
of the material is accelerated to these high $\g$ values.  The peak in
$\g$ leads to the highest observed temperature coming from the front
of the fireball and to lower temperatures coming from the interior
(Fig. 1.b).

In Fig. 2 we compare the calculated pulse at $t=153.7$ to a pulse
extrapolated using eqs. (10)--(11) from $t=14$. The agreement is very
good considering that the energy and matter density have decreased by
four and three orders of magnitude, respectively, while the maximal
Lorentz factor has increased by a factor of six between these two
moments.  The agreement is excellent in the inner region of the pulse
but is less satisfactory on the outside.  This is partly due to
continued acceleration of the outer regions where the pressure
gradient is steepest, and partly because of decreased numerical
accuracy in the regions where the density is very low.

Fig. 1.b shows profiles of $T_{obs}$, the temperature that would be
observed at infinity if the radiation could escape.  At early times,
there is a drop in $T_{obs}$ due to the broadening of the pulse as a
result of internal rearrangement.  Then, during the subsequent
radiation dominated phase, $T \propto \g^{-1}$ (eq 8), and the
observed temperature $T_{obs}$ of each shell remains a constant. The
overall spectrum of a fireball that becomes optically thin during this
phase is a blending of thermal spectra with different temperatures and
different blueshifts, and will be slightly broader than a single
temperature thermal spectrum (see Goodman, 1986).  The spectrum does
not change during the radiation dominated phase, apart from a minor
effect due to the addition of a harder component from the acceleration
of the outermost layers of the fireball. When a given shell enters the
matter dominated phase $T_{obs}$ begins to decrease, since $T$
continues to decrease but now without a compensating increase in $\g$
(eq 9).  The result is a softer spectrum than that observed during the
radiation dominated phase.  The emitted spectrum depends now on the
moment that each shell becomes optically thin.  Since different shells
become optically thin at different values of $\g$ we expect the
spectrum to be borader than that emitted during the radiation
dominated phase.

The evolution of a relativistic radiation fireball that we have
described here and in \S 2 is remarkably different from that of a
Newtonian fireball with $\eta_i \ll 1$.  Fig. 3 shows the energy
density and the mass density for a pulse with $\eta_i = 0.001$.  None
of the features described earlier appear.  Relativistic velocities are
never reached and the shape of the pulse is not frozen.  Instead we
observe an expanding, almost homogeneous sphere, rather than an
expanding shell of matter and radiation, and the expansion velocity of
most of the fireball is roughly the Newtonian velocity $\sqrt {2 \e_i}
= 0.045$.  A negligible fraction of the matter on the surface is
accelerated to higher speeds.  Interestingly, the Newtonian fireball
bears a strong qualitative resemblance to the relativistic fireball in
the local frame.  Therefore, the differences between the two cases
arise mainly because of the transformation to the observer frame.  In
the Newtonian case there is no difference between the observer frame
and the matter local frame, but in the relativistic case Lorentz
contraction leads to a drastic change in the appearance of the
fireball. Therefore, in the former case the fireball appears to fill
the entire sphere of radius $r$ whereas in the latter case the
observer sees a narrow pulse whose width remains of the same order as
the original width, leading to a time scale $\sim R_i/c$.

\bigskip
\centerline{\bf 4. Conclusions}
\bigskip

We have shown in this paper that fireballs with a large initial ratio
$\e_i$ of radiation energy to rest mass energy show certain common global
features during their expansion and evolution.  After a short initial
acceleration phase, the fluid reaches relativistic velocities, and the
energy and mass become concentrated in a radial pulse whose shape
remains frozen in the subsequent expansion.  The motion is then
described by an asymptotic solution (eqs 10, 11, \S 2), which gives
for each individual shell scaling laws similar to those of a
homogeneous sphere.

The expanding fireball has two basic phases: a radiation dominated
phase and a matter dominated phase.  Initially, during the radiation
dominated phase the fluid accelerates with $\g \propto r$ for each
Lagrangian shell.  The fireball is roughly homogeneous in its local
rest frame but due to the Lorentz contraction its width in the
observer frame is $\Delta r \approx R_i$, the initial size of the
fireball.  When the mean Lorentz factor of the fireball becomes $\bar
\g \approx (\eta_i+1) /2$ a transition takes place to the matter
dominated phase.  Ultimately, all the energy becomes concentrated in
the kinetic energy of the matter, and the matter coasts asymptotically
with a final Lorentz factor $\bar \g_f \approx (\eta_i+1)$.  The
matter dominated phase is itself further divided into two sub-phases.
At first, there is a frozen-coasting phase in which the fireball
expands as a shell of fixed radial width in its own local frame, with
a width $\sim \bar\gamma_f R_i \sim \eta_i R_i$.  Because of Lorentz
contraction the pulse appears to an observer with a width $\Delta r
\approx R_i$.  Eventually, the spread in $\bar\g_f$ as a function of
radius within the fireball results in a spreading of the pulse and the
fireball enters the coasting-expanding phase. In this final phase,
$\Delta r \approx R_it/\bar \g_f^2$, and the observed pulse width
increases linearly with time.

The fireball can become optically thin in any of the above phases.
Once this happen the system ceases to behave like a fluid, and the
radiation moves as a pulse with a constant width, while the baryons
enter a coasting phase like the one described above.

We have verified many of these theoretical results by means of
numerical simulations of spherically symmetric relativistic
fireballs (\S 3).  In particular, we confirm that the asymptotic
solution with frozen pulse shape is reproduced to a good
approximation.  This is a very useful result since it implies that in
future it is not necessary to carry out numerical simulations to very
late times.  As soon as the Lorentz factor of the expanding fluid
reaches a moderately large value, say $\bar\g \sim 10$, we can use the
theoretical results to extrapolate the pulse.  This will provide a
huge saving in computation, particularly in cases where $\e_i \gg 1$
and the asymptotic $\bar\g$ is very large.

An important aspect of fireball evolution that can be studied only by
numerical simulations is the early stages of rearrangement.  During
this phase the fireball is still only mildly relativistic and it
internally modifies the profiles of energy and matter density.  From a
number of simulations with different initial conditions, we find that
the ratio of energy to matter density, $\e$, usually ends up with a
lower value in the interior of the fireball and with large values on
the outside.  The Lorentz factor $\g$ also invariably increases from
the inside out.  These modified profiles enter the frozen pulse phase
and then do not change any further.  Consequently, it appears to be a
generic feature that any radiation that escapes from the fireball will
be hot on the outside and cooler on the inside.  In other words, the
observed radiation pulse will tend always to have a spectral profile
showing a characteristic hard-to-soft transition as a function of
time.  This effect will be enhanced if the early radiation from the
outside is emitted in the radiation-dominated phase and the later
radiation from the interior is released from matter-dominated layers.
The hard-to-soft signature will be even stronger in this case.  Even
if the radiation is not obtained directly from the fireball, but
through shock re-radiation as the fireball interacts with external
matter, this feature should still be present.

The discussion in this paper has been restricted to general issues
related to the evolution of relativistic fireballs.  The most
immediate application of these results is to cosmological and Galactic
halo models of gamma-ray bursts.  Although these models differ in
their explanations of the origin of the gamma-rays, all of them
involve a stage where the initially injected energy goes through a
fireball phase.  Therefore, the scaling laws that we have written down
for the matter and energy density, the temperature, and the Lorentz
factor $\g$ will be relevant.  Also, the hard-to-soft spectral
evolution described above should be observed in each sub-burst and
possibly across the whole burst as well.  In fact, this prediction is
probably valid regardless of how the final observed radiation is
produced, whether it be through direct emission from the fireball when
it becomes optically thin (Goodman 1986, Paczy\'nski 1986, Shemi \&
Piran 1990) or through shock re-emission (M\'es\'zaros \& Rees
1992a,b).

This work was supported by NASA grant NAG 5-1904 to Harvard University
and by a BSF grant to the Hebrew University..
\vfill\eject
\def\ppp{\par \medskip \noindent \hangindent .5in \hangafter 1}
\centerline{\bf References}
\bigskip
\ppp
Bowers R., L. \& Wilson J. R., 1991. {\it Numerical Modeling in Applied Physics
and Astrophysics}, Jones and Bartlett, Boston.
\ppp
Cavallo, G. \& Rees, M. J., 1978.
{\it Mon. Not. R. astr. Soc.}, {\bf 183}, 359.
\ppp
Goodman, J., 1986. {\it Astrophys. J. Lett.}, {\bf 308}, L47.
\ppp
M\'es\'zaros, P. \& Rees, M. J., 1992a.
{\it Mon. Not. R. astr. Soc.}, {\bf 258}, 41p.
\ppp
M\'es\'zaros, P. \& Rees, M. J., 1992b. {\it Astrophys. J. Lett.}, in press.
\ppp
Narayan, R., Paczy\'nski, B. \& Piran, T., 1992.
{\it Astrophys. J. Lett.}, {\bf 395}, L83.
\ppp
Paczy\'nski, B., 1986. {\it Astrophys. J. Lett.}, {\bf 308}, L51.
\ppp
Paczy\'nski, B., 1990. {\it Astrophys. J.}, {\bf 363}, 218.
\ppp
Piran, T. \& Shemi, A., 1993. {\it Astrophys. J. Lett.}, in press.
\ppp
Shemi, A. \& Piran, T., 1990. {\it Astrophys. J. Lett.},  {\bf 365}, L55.
\ppp
Vitello, P. \& Salvati, M. 1976. {\it Phys. of Fluids}, {\bf 19}, 1523.
\ppp
Weinberg, S., 1973. {\it Gravitation and Cosmology}, Wiley, New York.

\vfill\eject
\centerline{\bf Figure Captions}
\item{Fig. 1:} (a) Energy density,  $e \gamma$
(solid lines), and mass density, $n \g $ (dotted lines), in the
observer frame for a numerical simulation where the initial energy to
mass ratio $\e_i = 50$.  b. Lorentz factor, $\g$ (solid lines), and
observed temperature, $T_{obs} =\g T= \g e^{1/4} $ (dotted lines). The
temperature scale is arbitrary.

\item{Fig 2:}
Calculated (solid lines) and extrapolated (dotted lines) $n$, $e$ and
$\g$ profiles at $t=153.7$ (the end of the computation). The
extrapolation is from $t=14$ (shortly after the end of the
rearrangement phase) using eqs. (10)--(11).  The agreement is best in
the trailing edge of the pulse in the interior of the fireball and is
less satisfactory in the leading edge. This is because of the combined
effects of the steep pressure gradient and the loss of numerical
accuracy at lower densities.

\item{Fig. 3:}  Energy density, $e$
(solid lines), and mass density, $n$ (dotted lines), for a Newtonian
or non-relativistic fireball with $\e_i = 0.001$.  Note that, while
the final time is the same as in Fig. 1, the pulse has propagated a
much shorter distance. There is no hint of a shell structure or a
frozen pulse shape here, in contrast to the relativistic case shown in
Fig. 1.

\end